\documentclass[prl,twocolumn,amsmath,amssymb,showpacs,superscriptaddress]{revtex4}

\usepackage{graphicx}
\usepackage{dcolumn}
\usepackage{bm}
\usepackage{psfrag}

\def\be{\begin{equation}}
\def\ee{\end{equation}}
\def\ba{\begin{eqnarray}}
\def\ea{\end{eqnarray}}

\newcommand{\ie}{\emph{i.e.\ }}
\newcommand{\eg}{\emph{e.g.\ }}

\newcommand{\ket}[1]{|#1 \rangle}
\newcommand{\bra}[1]{\langle #1|}

\newcommand{\Ham}{\mathcal{H}}
\newcommand{\energy}{\mathcal{E}}

\begin{document}


\title{Tree-Level Electron-Photon Interactions in Graphene}

\author{Matthew Mecklenburg}
\email{meck0005@physics.ucla.edu}
\affiliation{Department of Physics and Astronomy, University of California, Los Angeles, California, 90095}
\affiliation{California NanoSystems Institute, University of California, Los Angeles, California, 90095}

\author{Jason Woo}
\affiliation{California NanoSystems Institute, University of California, Los Angeles, California, 90095}
\affiliation{Electrical Engineering Department,University of California, Los Angeles, California, 90095}

\author{B.C. Regan}
\email{regan@physics.ucla.edu}
\affiliation{Department of Physics and Astronomy, University of California, Los Angeles, California, 90095}
\affiliation{California NanoSystems Institute, University of California, Los Angeles, California, 90095}

%

\date{\today}

\begin{abstract}
Graphene's low-energy electronic excitations obey a 2+1 dimensional Dirac Hamiltonian.  After extending this Hamiltonian to include interactions with a quantized electromagnetic field, we calculate the amplitude associated with the simplest, tree-level Feynman diagram: the vertex connecting a photon with two electrons.  This amplitude leads to analytic expressions for the 3D angular dependence of photon emission, the photon-mediated electron-hole recombination rate, and corrections to graphene's opacity $\pi \alpha$ and dynamic conductivity $\pi e^2/2 h$ for situations away from thermal equilibrium, as would occur in a graphene laser. We find that Ohmic dissipation in perfect graphene can be attributed to spontaneous emission.

\end{abstract}

\maketitle

\begin{figure}\begin{center}
	\includegraphics[width=3.5in]{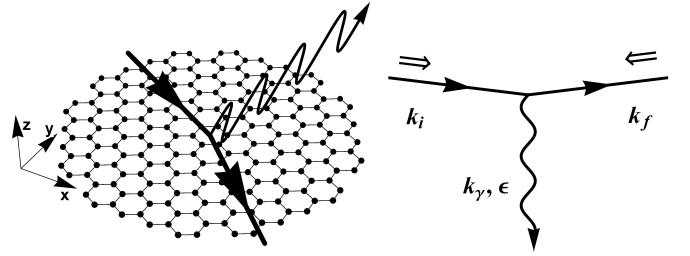}
	\caption{\label{fig:feynman} Schematic drawing of a representative emission process (left), and the corresponding Feynman diagram (right). The photon lives in 3D space, while the electrons are confined to the graphene sheet. The initial electron is described by its momentum $\mathbf{p}_i=\hbar \mathbf{k}_i$ and its pseudospin $\Rightarrow$, which for a conduction electron near $\mathbf{K}^+$ is directed along $\mathbf{k}_i$. Interacting with the photon (wavevector $\mathbf{k}_\gamma$ and polarization $\boldsymbol{\varepsilon}$) destroys the conduction electron, creating a valence band electron with momentum $\hbar \mathbf{k}_f$ and pseudospin $\Leftarrow$.}
\end{center}\end{figure}

Electron-photon interactions determine the opto-electronic properties of a material.  The electrons in graphene, a single atomic layer of graphite, exhibit superlative electronic properties associated with their exotic Hamiltonian \cite{2009NetoReview,2010DasSarma}.  In particular, a tight binding model \cite{1947Wallace} of graphene produces a Hamiltonian that, for low energy excitations, is formally identical to a 2+1 dimensional Dirac equation for massless fermions \cite{1984Semenoff}, with the Fermi velocity and the sublattice state vector filling the roles of the speed of light and spin respectively.  As part of an effort to understand how electron-hole recombination might limit the function of a graphene-based transistor, we use this Dirac Hamiltonian to calculate the amplitude for the electron-photon interaction diagrammed in Fig.~\ref{fig:feynman}.  Rotating this diagram with respect to the time axis allows the consideration of both photon emission (\ie recombination) and absorption rates, which we relate to graphene's opacity and dynamic conductivity.  

These measurable \cite{2005Novoselov,2008Nair,2008Mak,2008Kuzmenko} properties have been previously treated using semiclassical methods (where the electromagnetic field is not quantized) within the Kubo and Landauer formalisms \cite{2002Ando,2006GusyninPRL,2007Ryu,2007Ziegler,2009Vildanov} and perturbation theory \cite{2008Nair, 2009Lewkowicz}. Our fully quantum mechanical calculation reproduces results found previously, such as $\pi \alpha$ for the optical opacity \cite{2008Nair,2008Kuzmenko,2008Stauber} and $\pi e^2/2 h$ \cite{2006GusyninPRL,2007Ziegler,2007Ryu,2008Stauber,2009Vildanov, 2009Lewkowicz} for the zero-temperature conductivity. We extend these previous results to non-equilibrium situations (\eg population inversion) and specify the full angular dependence of photon emission/absorption. Furthermore, we identify spontaneous emission as the mechanism of dissipation, present even in idealized graphene, that is usually left unspecified \cite{2007Ziegler,2008Stauber,2009Vildanov,2009Lewkowicz}.  

\begin{figure}\begin{center}
	\includegraphics[width=3.25in]{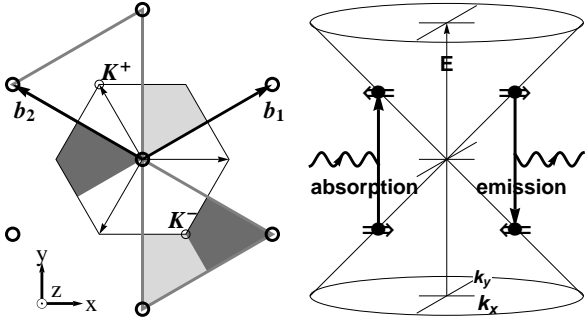}
	\caption{\label{fig:geo} The hexagonal first Brillouin zone (left) and the dispersion relation near the points $\mathbf{K}^\pm$ (right).  On the left, the $\mathbf{K}^+$ points are indicated by thin arrows, and the reciprocal lattice primitive vectors $\mathbf{b}_i$ by thick arrows.  Shading indicates how translating some slices of the hexagon by reciprocal lattice vectors reconstructs an equivalent Brillouin zone, here shown in a bowtie configuration, that centers the inequivalent $\mathbf{K}^\pm$ points in two triangular regions.  Near the $\mathbf{K}^\pm$ points the dispersion relation is linear in $\mathbf{|k|}$, which gives the Dirac cones shown on the right.  Absorption or emission of a photon transfers an electron from one cone to the other.}
\end{center}\end{figure}

The carbon atoms in graphene form a two-dimensional honeycomb network with two inequivalent atomic sites per unit cell.  In the simplest tight-binding description of graphene, an electronic energy $\energy$ is associated with each atomic site in the sheet, and an energy $t$ parametrizes the probability of an electron hopping from one site to its neighbor on the other sublattice.  An operator $A^\dagger_{\mathbf{R}_j}$ creates a $2P_z$ electron on the `A' site in cell $j$, with a corresponding destruction operator $A_{\mathbf{R}_j}$. With similar operators for the `B' sites, the total Hamiltonian $H$ is
\begin{equation}\label{eq:originalTB}
H =  \energy\sum_{j}  (A_{\mathbf{R_j}}^{\dagger} A_{\mathbf{R_j}}^{\phantom{\dagger}} +   B^{\dagger}_{\mathbf{R_j}} B^{\phantom{\dagger}}_{\mathbf{R_j}})
-t \sum_{<i,j>} (A^{\dagger}_{\mathbf{R_i}} B^{\phantom{\dagger}}_{\mathbf{R_j}} + \textrm{h.c.}),
\end{equation}
where $j$ runs over the $N$ sites in the sheet, and $i$ runs over the nearest neighbors of the site $j$. Spin indices on the operators and the sums are understood. Fourier transforming the creation and annihilation operators (\eg $A_{\mathbf{R}_i}=\sum_{j} A_{\mathbf{Q}_j} \exp(i \mathbf{R}_i\cdot\mathbf{Q}_j)/\sqrt{N}$, where the $\mathbf{Q}_j=\frac{m}{N_1} \mathbf{b}_1+\frac{n}{N_2} \mathbf{b}_2$ are the $N=N_1 N_2$ wavevectors in the first Brillouin zone) allows the Hamiltonian (\ref{eq:originalTB}) to be written,
\begin{equation}\label{eq:fourierTB}
H =  \sum_{j} \begin{pmatrix} A^{\dagger}_{\mathbf{Q}_j} & B^{\dagger}_{\mathbf{Q}_j} \end{pmatrix} \Ham
\begin{pmatrix} A^{\phantom{\dagger}}_{\mathbf{Q}_j} \\ B^{\phantom{\dagger}}_{\mathbf{Q}_j} \end{pmatrix}.
\end{equation}
There are two spin states per $\mathbf{Q}_j$, and two mobile $2P_z$ electrons per cell, so the first Brillouin zone (Fig.~\ref{fig:geo}) is exactly filled in electrically neutral graphene at zero temperature. The energy origin is set at the energy of the highest occuppied states, which are those at the Brillouin zone corners $\mathbf{K}^\kappa=\kappa \frac{2 \mathbf{b}_2+\mathbf{b}_1}{3}+m \mathbf{b}_1+n \mathbf{b}_2$ \cite{2009Bena}.  The label $\kappa=\pm 1$ indexes the two inequivalent corners.   For $\mathbf{Q}$ near a $\mathbf{K}^\kappa$ point the single-particle Hamiltonian $\Ham$ is
\begin{equation}\label{eq:HamP}
\Ham=v_F(\kappa \sigma_x p_x +  \sigma_y p_y),
\end{equation}
where the momentum $\mathbf{p}=\hbar \mathbf{k}=\hbar( \mathbf{Q}-\mathbf{K})$.  With  $\phi$ defined by the in-plane components $\vec{p}$ of $\mathbf{p}=\vec{p}+p_z \mathbf{\hat{z}}$ according to $\vec{p}= p (\cos\phi \mathbf{\hat{x}}+\sin\phi \mathbf{\hat{y}})$, the corresponding eigenvalue equation is
\begin{equation}\label{eq:PhiMatrix}
\Ham \ket{\chi}=\frac{\kappa v_F p}{\sqrt{2}}\begin{pmatrix}
0 & e^{-i \kappa \phi} \\
 e^{i \kappa \phi} & 0
\end{pmatrix}
\begin{pmatrix}
e^{-i \kappa \phi/2} \\
\beta \kappa e^{i \kappa \phi/2} 
\end{pmatrix}=\beta v_F p \ket{\chi},
\end{equation}
where the band index $\beta = \pm 1$ labels whether the energy is positive or negative (\ie conduction or valence).  Thus the Hamiltonian (\ref{eq:HamP}) produces a linear dispersion relation $\energy=\pm v_F p$.  The product $\beta \kappa/2$ gives the helicity eigenvalue for the state $\ket{\chi}$, where the helicity operator is defined as $\hat{h}= \Ham/(2 \kappa v_F p)$.

Having identified the eigenspinors $\ket{\chi(\vec{p},\beta,\kappa)}$ of the single particle Hamiltonian $\Ham$, we can re-write the total Hamiltonian $H$ in terms of operators that create ($C^\dagger_\mathbf{Q}$) and destroy ($C_\mathbf{Q}$) energy eigenstates, 
\begin{equation}
\begin{split}\label{eq:CMatrix}
H=\sum_{\mathbf{Q}}&\kappa v_F p (\bra{\chi_c} C^\dagger_{c,\mathbf{Q}}+\bra{\chi_v} C^\dagger_{v,\mathbf{Q}})\\
&\times \begin{pmatrix}
0 & e^{-i \kappa \phi} \\
 e^{i \kappa \phi} & 0
\end{pmatrix}
(\ket{\chi_c} C_{c,\mathbf{Q}}+\ket{\chi_v} C_{v,\mathbf{Q}})
\end{split}
\end{equation}
where the sum is over $\mathbf{Q}$ near $\mathbf{K}^\kappa$ and $c$ ($v$) refers to the conduction (valence) band.
 
We introduce the electromagnetic field with a Peierls (minimal coupling) substitution  $\mathbf{p}\rightarrow \mathbf{p}-q \mathbf{A}/ c$, treating the new vector potential term \cite{1967Sakurai} as a quantized perturbation $H'$ in the full Hamiltonian $H=H_0 + H'$,
\begin{equation}
\mathbf{A}(\textbf{r},t) = c \sum_{\mathbf{k}j} \sqrt{\frac{2 \pi \hbar}{\epsilon_r V \omega}}%
 (\hat{\boldsymbol{\varepsilon}}_{j} C_{\mathbf{k}j} e^{i \mathbf{k} \cdot \mathbf{r}-i\omega t}  + \hat{\boldsymbol{\varepsilon}}_{j}^{*} C^{\dag}_{\mathbf{k}j} e^{-i \mathbf{k} \cdot \mathbf{r}+i\omega t })
\label{vecpot}
\end{equation}
Here $j$ indexes the photon's polarization states,  $V$ is the normalization volume, $\epsilon_r$ is the relative permittivity, and $\omega=c|\mathbf{k}|$.  As is evident from the appearance of the speed of light $c$ (and not the Fermi velocity $v_F$) in this substitution, the electron-photon coupling implied follows from the local gauge invariance of the standard model Lagrangian, and is not related to the properties of the Hamiltonian (\ref{eq:HamP}) under gauge transformations.  

The electron-photon interaction rate can be calculated using the standard arguments of Fermi's Golden Rule, suitably modified to account for the system's mixed dimensionality.  The rate $\Gamma_{i\rightarrow f}$ to go from an eigenstate $\ket{\varphi_i}$ of the unperturbed electronic Hamiltonian $H_0$ to a given final state $\ket{\varphi_f}$ is 
\begin{align}
\Gamma_{i\rightarrow f}&=\frac{d}{dt}|\langle{\varphi_f (t)}|{\psi(t)}\rangle|^2\,\text{, where}\\
\langle{\varphi_f (t)}|{\psi(t)}\rangle&=\frac{1}{i \hbar}\int_0^{t} \bra{\varphi_f(t')}\Ham'(t')\ket{\varphi_i(t')}dt'\equiv M.
\end{align}
In the position representation, the time-dependent solutions to the unperturbed electronic $\Ham$ have the form
\begin{equation}
\bra{\vec{r}}\varphi(t)\rangle=\frac{1}{\sqrt{A}}e^{i(\vec{k}\cdot \vec{r}-\omega t)} F(z) \ket{\chi(\hbar\vec{k},\beta,\kappa)},
\end{equation}
where $\omega=v_F |\vec{k}|$ and $A$ is the graphene area.  Initially we consider processes that create a valence electron  $\ket{\varphi_f}\propto\ket{\chi (\hbar \vec{k}_v,-1,\kappa)}$ and a photon $\ket{\mathbf{k}_\gamma,\hat{\boldsymbol{\varepsilon}}}$, while destroying a conduction electron $\ket{\varphi_i}\propto\ket{\chi (\hbar\vec{k}_c,1,\kappa)}$.  Then
\begin{align}
M&=\frac{i q v_F}{\hbar}\sqrt{\frac{2 \pi \hbar}{V \omega \epsilon_r}} \int_0^{\Delta t} e^{i(\omega_v + \omega - \omega_c)t'}dt' \label{eq:M}\\
&\quad \times \int_A e^{-i(\vec{k}_v + \vec{k}_\gamma-\vec{k}_c)\cdot\vec{x}'}\frac{d^2x'}{A}\int |F(z')|^2 e^{-i k_{\gamma z} z'}dz'\nonumber\\
&\quad \times \bra{\chi_v} \vec{\sigma}\cdot \hat{\boldsymbol{\varepsilon}}^*\ket{\chi_c}
\bra{n'_c}\bra{n'_v}\bra{n'_\gamma}C_v^\dagger C_\gamma^\dagger C_c \ket{n_\gamma}\ket{n_v}\ket{n_c}.\nonumber
\end{align}
Derived from the $z$-extent of the carbon $2P_z$ atomic orbitals, the normalized function $F(z)$ is only appreciable within a few angstroms of the graphene plane.  Since we are considering photons with optical or longer wavelengths $\lambda=2 \pi/k_\gamma$, the integral over $dz'$ gives unity to excellent approximation.  In atomic physics this step applies to all three spatial dimensions ($e^{i \mathbf{k}\cdot \mathbf{r}}\sim 1$) and is known as the dipole approximation.

We square $M$, and consider the interval $\Delta t$ to be short compared to the lifetime $1/\Gamma$ and long compared to the time scale $1/\omega$ set by the energy of the transition, \ie $1/\Gamma\gg\Delta t\gg 1/\omega$.  In this limit, with large area $A$,
\begin{align}
 d\Gamma_{c\rightarrow v}&=\frac{q^2 v_F^2}{\hbar}\frac{2 \pi \hbar}{A^2 V \omega \epsilon_r}|\bra{\chi_v} \vec{\sigma}\cdot \hat{\boldsymbol{\varepsilon}}^*\ket{\chi_c}|^2\nonumber\\
 &\quad \times 2 \pi \delta(\hbar(\omega_v+\omega-\omega_c))\,(2 \pi)^2 A \,\delta^2(\vec{k}_v + \vec{k}_\gamma - \vec{k}_c) 
  \nonumber \\
 &\quad \times n_c (1-n_v)(1+n_\gamma)\, \frac{A\,d^2\vec{k}_v }{(2 \pi)^2}\, \frac{V\, d^3\mathbf{k}_\gamma}{(2 \pi)^3}, \label{eq:RateResult}
\end{align}
where we have used the standard relations $C^\dagger\,\ket{n}=\sqrt{1\pm n}\,\ket{1\pm n}$, $C\,\ket{n}=\sqrt{n}\,\ket{\pm(n-1)}$, and $\bra{n'}n\rangle=\delta_{n'n}$, with the upper (lower) sign chosen for bosons (fermions). Thus we see that the recombination rate is proportional to the number of conduction electrons $n_c$ and the number of holes $1-n_v$.  The first and second parts of $1+n_\gamma$ correspond to spontaneous and stimulated emission respectively \cite{1964Lasher}.

To evaluate the angular matrix element in (\ref{eq:RateResult}), we define an orthonormal triple $\hat{\mathbf{k}}_\gamma=(\sin{\theta_{\gamma}} \cos{\phi_{\gamma}}, \sin{\theta_{\gamma}} \sin{\phi_{\gamma}}, \cos{\theta_{\gamma}})$, $\hat{\boldsymbol{\varepsilon}}_1=\hat{\mathbf{z}}\times \hat{\mathbf{k}}_\gamma/|\hat{\mathbf{z}}\times \hat{\mathbf{k}}_\gamma|$, and $\hat{\boldsymbol{\varepsilon}}_2=\hat{\mathbf{k}}_\gamma\times \hat{\boldsymbol{\varepsilon}}_1/|\hat{\mathbf{k}}_\gamma\times \hat{\boldsymbol{\varepsilon}}_1|$ that describes the photon and its possible polarizations.  Summing over the possible polarizations $j$ of the created photon gives
\begin{equation}\label{eq:angle}
\sum_{j}|\bra{\chi_v} \vec{\sigma}\cdot \hat{\boldsymbol{\varepsilon}}_j^*\ket{\chi_c}|^2
= 1-\sin^2\theta_\gamma \sin^2(\phi_c/2+\phi_v/2-\phi_\gamma).
\end{equation}
As $p_z$ does not appear in $\Ham$, the component of the photon polarization along $\hat{\mathbf{z}}$ does not contribute to this matrix element.  The integrals over the energy and momentum $\delta$-functions in (\ref{eq:RateResult}) can now be performed, with the result  
\begin{align}
 &\frac{d\Gamma_{c\rightarrow v}}{d\Omega_\gamma}=\frac{q^2}{\hbar c} \left(\frac{v_F}{c}\right)^2 \frac{\omega_c}{\pi \epsilon_r}\,n_c (1-n_v)(1+n_\gamma) \label{eq:RateResult2}\\
 &\quad \times \frac{1 -  \frac{2v_F}{c} \sin\theta_\gamma \cos(\phi_\gamma-\phi_c) + (\frac{v_F}{c})^2 \sin^2\theta_\gamma}{(1-(\frac{v_F}{c})^2 \sin^2\theta_\gamma)^2}
  \nonumber \\
 &\times \left(1-\frac{\sin^2\theta_\gamma \sin^2(\phi_\gamma-\phi_c)}{1 -  \frac{2v_F}{c} \sin\theta_\gamma \cos(\phi_\gamma-\phi_c) + (\frac{v_F}{c})^2 \sin^2\theta_\gamma}\right). \nonumber
\end{align}
The last line in (\ref{eq:RateResult2}) corresponds to the angular matrix element (\ref{eq:angle}).

\begin{figure}\begin{center}
	\includegraphics[width=3in]{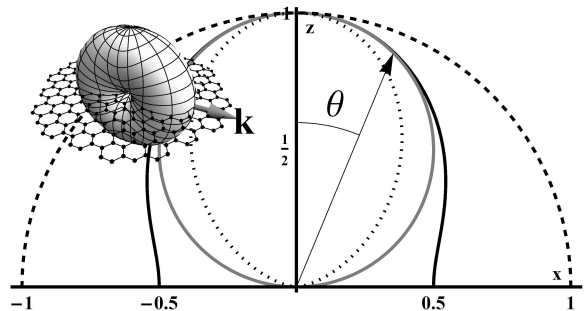}
	\caption{\label{fig:2Dangle} Polar plot of photon emission distributions in the $xz$ plane.  The dashed (dotted) curve corresponds to the emission from an initial electron moving along the $\mathbf{x}$-axis ($\mathbf{y}$-axis), while the solid black curve represents the average over all $\vec{k_i}$ directions. The Lambertian function $\cos \theta$ is shown in grey for comparison. The inset shows the 3D pattern for one  choice of $\mathbf{k}$. An electron moving along $\mathbf{y}$ emits a $\mathbf{x}$-polarized photon, and thus cannot emit in the $\mathbf{x}$ direction.}
\end{center}\end{figure}

Since $v_F/c$ is a small number $\sim 1/300$ \cite{2005Novoselov}, several approximations are in order. To better than 1\% accuracy  $\vec{k}_c\simeq \vec{k_v}$ and $k_\gamma/k_{c(v)}\simeq 2 v_F/c$.  The energy of the initial conduction electron is half that of the photon, and of the same magnitude but opposite sign of the final valence electron.  The photon's momentum is negligible in comparison to the electrons'; as a result $\phi_c\approx \phi_v$ and $\mathbf{K}^+ \leftrightarrow \mathbf{K}^-$ transitions are impossible in this low energy limit.  To lowest order in $v_F/c$ the angular dependence of  (\ref{eq:RateResult2}) is $(1 -\sin^2\theta_\gamma \sin^2(\phi_\gamma-\phi_c)- 2 \frac{v_F}{c} \sin\theta_\gamma \cos(\phi_\gamma-\phi_c))$.  Thus for small $v_F/c$ a conduction electron is slightly more likely to emit a photon opposite $\vec{k}_c$ than along $\vec{k}_c$. Figure~\ref{fig:2Dangle} shows various plots of the angular distribution in the small $v_F/c$ limit, which we will adopt henceforth. 

When averaged over the possible momentum directions of the conduction electron, the emission or absorption of a photon depends on the polar angle $\theta_\gamma$ from the normal to the graphene sheet like $1-\frac{1}{2}\sin^2\theta_\gamma$. Because this function falls off more slowly than the Lambertian function $\cos \theta$, a graphene sheet will appear progressively brighter (\ie blacker) at angles away from normal incidence.  At this level of analysis the angular matrix element (\ref{eq:angle}), and thus the rate, is zero for the metallic nanotube case \cite{2004Jiang,2009spin}.  The interaction Hamiltonian contains only photons polarized along the nanotube axis, and such photons do not couple the initial and final electronic states.

The form of the matrix element (\ref{eq:angle}) indicates that angular momentum conservation is enforced in an unusual way in this problem. In a more conventional condensed matter system a typical optical transition involves bands with different orbital angular momentum quantum numbers, and allows the possibility of a spin flip.  For instance, in gallium arsenide interband transitions occur between orbitals with $S$ and $P$ symmetries \cite{2004Zutic}. Here the transition is $2P_z\rightarrow 2P_z$ and there is no spin flip. Thus the usual sources of angular momentum for the photon do not contribute in graphene.  The structure of the matrix element (\ref{eq:angle}), which follows directly from the Hamiltonian (\ref{eq:HamP}) and the assumption of minimal coupling, implies that the pseudospin flip creates the angular momentum $\hbar$ of the photon.  We further explore this connection between pseudospin and angular momentum elsewhere \cite{2009spin}. 

For states connected by the $\delta$-functions in Eq.~(\ref{eq:RateResult}), the proportionality $\Gamma_{c\rightarrow v}\propto n_c\left(1-n_v\right)\left(1+n_\gamma\right)$ is general, and applies whether the $n$'s reflect equilibrium distributions or not.  Non-thermal distributions are commonly handled by introducing a quasi-Fermi level that differs for electrons and holes \cite{1952Shockley}. To simply illustrate the time scales relevant for photon-mediated electron-hole recombination, we consider a perfect population inversion, \ie  $n_c=1$ and $n_v=0$. Integrating (\ref{eq:RateResult2}) over all directions of $\mathbf{k}_\gamma$ gives the rate for a conduction electron with energy $\energy_c= \hbar \omega_c=v_F\hbar k_c$ to decay spontaneously ($n_\gamma=0$) via photon emission, 
\begin{equation}\label{eq:lifetime}
\Gamma_{c\rightarrow v}= \frac{8 \alpha}{3\epsilon_r} \left(\frac{v_F}{c}\right)^2 \omega_c,
\end{equation}
where $\alpha=e^2/\hbar c\simeq 1/137$ is the fine structure constant.  This rate corresponds to a lifetime $\tau=1/\Gamma$ of about 3~ns for a 1~$e$V conduction electron.
  
For thermal populations the averaged transition matrix element $\langle |M|^2 \rangle =\text{Tr} \{ \rho_c \rho_v \rho_\gamma |M|^2 \}$, where the density operators $\rho$ are given by $\rho=e^{-\Ham/kT}/\text{Tr} \{e^{-\Ham/kT}\}$ and the trace is taken over the possible occupations: $n\in [0,1]$ for the electrons, and $n \in [0, \infty]$ for the photons \cite{1977CohenT}.  Evaluating the trace gives Bose and Fermi distribution functions,
\begin{align}\label{eq:thermalrate}
\langle \Gamma_{c\rightarrow v}\rangle&=  \frac{8 \alpha\energy_c}{3\epsilon_r \hbar} \left(\frac{v_F}{c}\right)^2  \left(\frac{1}{e^{\frac{\energy_c-\mu}{kT}}+1}\right)\left(1-\frac{1}{e^{\frac{-\energy_c-\mu}{kT}}+1}\right) \nonumber\\&\quad \left(1+\frac{1}{e^{2\energy_c/kT}-1}\right),
\end{align}
where we have allowed for a chemical potential $\mu$. The second line of (\ref{eq:thermalrate}) shows that recombination stimulated by the blackbody background becomes important for $2 \energy_c=\hbar \omega \lesssim kT$.  At room temperature with $\mu=0$ a conduction state with energy $\energy_c=kT\simeq0.025$~$e$V will be populated and decay with a characteristic lifetime of about 400~ns. For many practical purposes this rate is negligible, since,  for instance, the second order (Auger) process gives picosecond lifetimes \cite{2007Rana}.

In contrast, the time-reverse of this recombination process, photon absorption, is observable practically by the unaided eye \cite{2008Kuzmenko,2008Nair}. To analyze absorption we proceed as in the derivation of Eqs.~(\ref{eq:RateResult2}--\ref{eq:lifetime}), this time considering illumination normally incident on the graphene plane ($\theta_\gamma=0$) at a rate $\Gamma_i= n_\gamma c A/\epsilon_r V$.  Then the promotion rate $\Gamma_{v\rightarrow c}$ from the valence to the conduction band is 
\begin{equation}
\Gamma_{v\rightarrow c}=\pi \alpha \,(1-n_c) \, n_v \Gamma_i,\label{eq:promotion}
\end{equation}
where we have included a factor of 4 for the valley and (normal) spin degeneracies.  Discounting spontaneous emission into the illuminating beam, we take the net absorption rate $\Gamma_{abs}$ to be the promotion rate minus the stimulated emission rate $\propto  n_c \,(1-n_v)\,n_\gamma$, which gives
\begin{equation}
\Gamma_{abs}=\pi \alpha \,(n_v-n_c) \Gamma_i.\label{eq:pialpha}
\end{equation}
With initial $n_c=0$ and $n_v=1$ Eq.~(\ref{eq:pialpha}) reproduces the $\pi \alpha$ result for the optical absorption of a graphene sheet \cite{2008Kuzmenko,2008Nair}, and  identifies spontaneous emission as the source of dissipation.  For $n_c>n_v$ the absorption is negative, implying gain and the possibility of a graphene laser \cite{1964Lasher,2007Ryzhii}.  As before, thermally averaging (\ref{eq:pialpha}) replaces the $n$'s with Fermi functions, with the result that the absorption goes to zero for $\hbar \omega \ll kT$ or $\hbar \omega \ll 2 |\mu|$.

We can relate the energy absorption rate implied by (\ref{eq:pialpha}) to the conductivity $\sigma$ by invoking Ohm's Law, which implies that the power dissipated per unit area is $\mathbf{K}\cdot\mathbf{E}=\sigma \mathbf{E}^2$. Here $\mathbf{K}$ is the current density and $\mathbf{E}=-\frac{1}{c}\partial \mathbf{A}/\partial t$ is the electric field.  Since the energy density of the electromagnetic field is $\epsilon_r\mathbf{E}^2/4 \pi= n_\gamma \hbar \omega/V$, we have 
\begin{equation}\label{eq:cond}
\sigma= \frac{\alpha c}{4}\,  (n_v-n_c)= \frac{\pi e^2}{2 h}\,(n_v-n_c),
\end{equation}
which is $\pi e^2/2 h$ at $T=0$. This expression can be written  
\begin{equation}\label{eq:thermalsigma}
\sigma = \frac{\pi e^2}{4 h}\left[ \tanh \left(\frac{\hbar \omega + 2 \mu}{4 k T}\right)
 +\tanh\left(\frac{\hbar \omega - 2 \mu}{4 k T}\right)\right].
\end{equation}
after thermal averaging, which is identical to the result found previously \cite{2008Stauber,2008Kuzmenko,2008Mak}. Our calculation, like the previous ones, is not rigorous at $\omega=0$, as the dc limit explicitly violates the assumption required to generate the energy $\delta$-function in Eq.~(\ref{eq:RateResult}).

In conclusion, we have performed the first calculation of graphene's optical properties with a quantized electromagnetic field.  The calculation is fully quantum mechanical and free of thermodynamic assumptions until the final step, which allows the treatment of systems far from thermal equilibrium.  Furthermore, the inherently three-dimensional formalism gives amplitudes as a function of photon polarization and propagation direction relative to the graphene plane.  The dependence on electron and photon state occupation numbers follows directly from their fermionic and bosonic commutation relations.  The former result in Pauli blocking, while the latter give terms that can be identified with spontaneous and stimulated emission.  Spontaneous emission proves to be a source of dissipation present even in idealized graphene, with an implied violation of time-reversal symmetry whose introduction can be traced back to the use of Fermi's Golden Rule.  Stimulated emission from graphene could prove technologically useful, since an electronic population inversion would allow graphene to serve as the gain medium in a laser tunable over a broad band of frequencies.

This work was supported in part by the DARPA CERA program.
%


\bibliography{graphene}

\end{document}